\documentclass[pra,aps,showpacs,showkeys,superscriptaddress,reprint,amsmath,amssymb]{revtex4-1}
\usepackage{graphicx}
\usepackage{color}
\usepackage{amssymb,amsmath,graphicx}
\usepackage{comment}
\usepackage{verbatim}
\usepackage{hyperref}
\hypersetup{
bookmarks=false,
pdftitle={Solitons in open & driven BH chains},
anchorcolor=blue,
colorlinks=true,
citecolor=blue,
urlcolor=blue,
linkcolor=blue}
\usepackage{marginnote}
\usepackage[author=smbdy]{pdfcomment}
\usepackage{xcolor}

\begin{document}

\title{Stationary discrete solitons in circuit QED }

\author{Uta Naether}\email{Corresponding author: naether@unizar.es}
\affiliation{Instituto de Ciencia de Materiales de Arag\'on y
  Departamento de F\'{\i}sica de la Materia Condensada,
  CSIC-Universidad de Zaragoza, Zaragoza, E-50012, Spain.}
\author{Fernando Quijandr\'{\i}a} 
\affiliation{Instituto de Ciencia de Materiales de Arag\'on y
  Departamento de F\'{\i}sica de la Materia Condensada,
  CSIC-Universidad de Zaragoza, Zaragoza, E-50012, Spain.}
\author{Juan Jos\'e Garc\'{\i}a-Ripoll}
\affiliation{Instituto de F\'{\i}sica Fundamental, IFF-CSIC, Serrano 113-bis, Madrid E-28006, Spain}
\author{David Zueco} 
\affiliation{Instituto de Ciencia de Materiales de Arag\'on y Departamento de F\'{\i}sica de la Materia Condensada, CSIC-Universidad de Zaragoza, Zaragoza, E-50012, Spain.}
\affiliation{Fundaci\'on ARAID, Paseo Mar\'{\i}a Agust\'{\i}n 36, Zaragoza 50004, Spain}

\date{\today}

\begin{abstract}
We demonstrate that  \emph{stationary} localized solutions (discrete solitons) exist in a one dimensional Bose-Hubbard   lattices with gain and loss in the semiclassical regime.
Stationary solutions, by definition, are robust and do not demand for state preparation. 
Losses, unavoidable in experiments, are not a drawback, but a necessary ingredient for these modes to exist.
The semiclassical calculations are complemented with their classical
limit and  dynamics based on a Gutzwiller  \emph{Ansatz}.
We argue that circuit QED architectures are ideal platforms for realizing the physics developed here.
Finally, within the input-output formalism, we explain how to experimentally access the different phases, including the solitons, of the chain.
\end{abstract}



\maketitle 

\section{Introduction}

Realizations of quantum nonlinear media as ultracold atoms in optical lattices\cite{Morsch2006}, ion-traps \cite{Blatt2012} or superconducting circuits \cite{Naether2014,Raftery2014}  are interesting candidates for future quantum information processors.   Apart from this challenging goal, they  are also testbeds to explore new many body states of matter both in the classical and quantum regime  \cite{Houck2012}.  Among others,  solitons \emph{- localized and form  preserving solutions -} are  a paradigmatic example of collective nonlinear solutions. In the so-called \emph{classical} limit for the Bose-Hubbard (BH) model, the operators are replaced by their c-number average, obtaining  the well known discrete nonlinear Schr\"odinger equation (DNLS) \cite{Kevrekidis}. Stable exact and numerical localized solutions, discrete solitons, exist in different dimensions and topologies \cite{Flach2008, Lederer2008, Kartashov2011,Naether2013}.

Quantum solitons have been hypothesized to exist in the BH model with and without dissipation. Theoretical predictions based on different approaches, 
Gutzwiller- \cite{Krutitsky2010} , truncated Wigner-approximations
\cite{Martin2010}, Gaussian expansions \cite{Witthaut2011} as well as  density matrix renormalization group
techniques \cite{Mishmash2009} have found slowly decaying
 localized solutions.  Therefore, none of those were stable
solutions for the dynamics. Thus,  quantum fluctuations seem to kill these topological
solutions. Experimental realizations in the quantum realm are few.  
 Bose-Einstein condensates confirmed the temporal existence of bright \cite{Burger1999} and dark \cite{Strecker2002} localized modes\cite{Frantzeskakis2010}. For ions in optical traps, a proposal for the observation of solitons \cite{Landa2010} was shortly after followed by the experimental observation \cite{Mielenz2013}.

Things \emph{may} change if  dissipation and gain  coexist.
In the classical limit yielding the dissipative driven
DNLS (DD-DNLS) equation, localized solutions  have
been reported  \cite{Peschel2004,
  Prilepsky2012}. Furthermore, the DD-DNLS exhibits "spontaneous
walking" solitons \cite{Egorov2013}; using nonlinear gain and
dissipation, exact travelling discrete solitons exist as stable
dynamical attractors \cite{Johansson2014}. 
Therefore, an \emph{open question remains} in the literature.  What about
quantum solitons in nonlinear media with loss and gain?  In our opinion, the combination of many body physics, dissipation and
driving is interesting.  It provides new phases to explore with non
thermal but equilibrated states, as already demonstrated in the
dissipative driven BH (DD-BH)
model \cite{Jin2013, Jin2014}.  Besides, it establishes a links with man made
realizations of lattice systems where dissipation can be an issue
\cite{Houck2012}.  
In the present context these novel phases could provide solitons.

\begin{figure}[b]
\centerline{\includegraphics[width=1.\columnwidth]{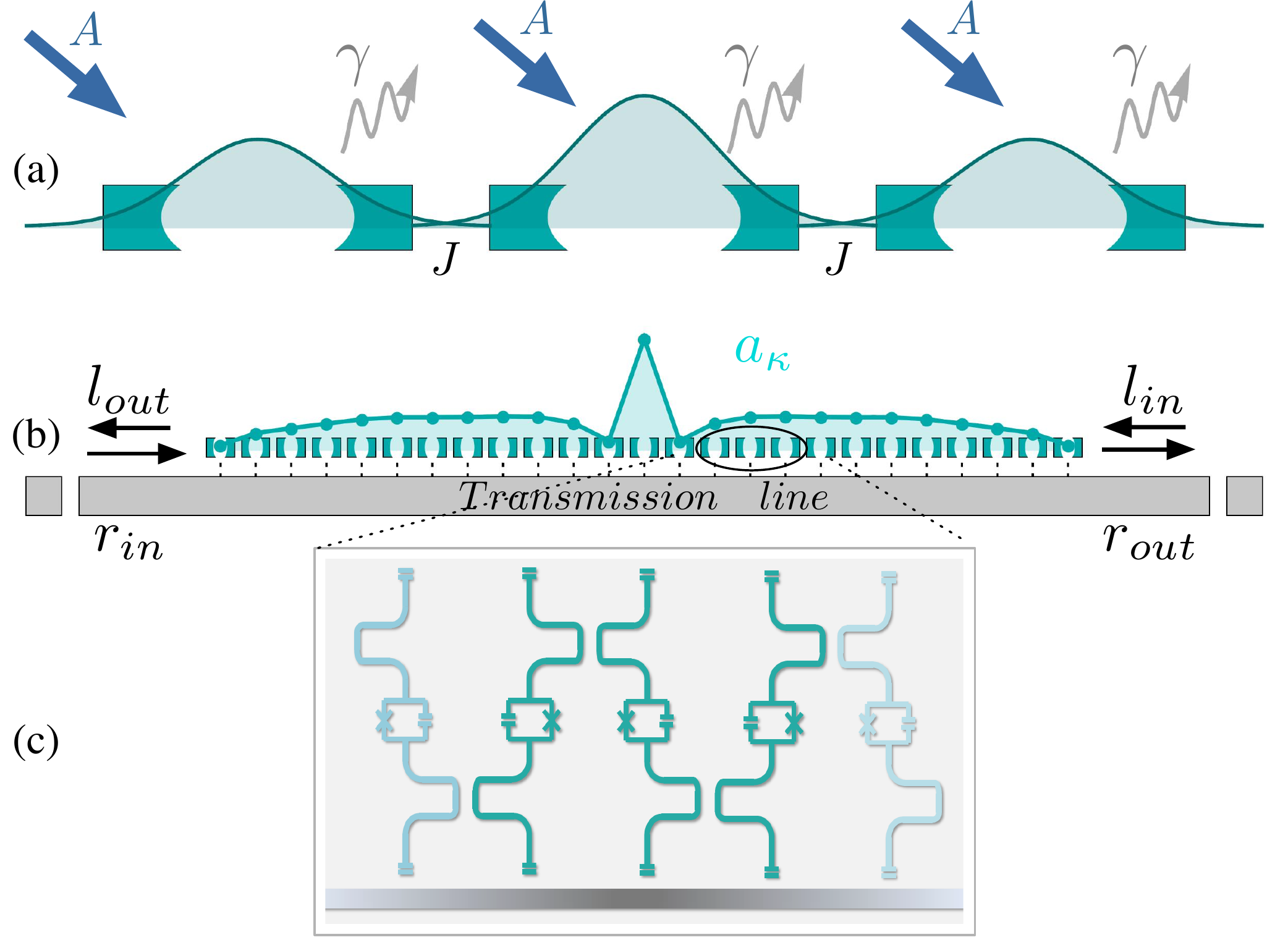}}
\caption{(Color online) (a) Scheme of the driven and dissipative cavity array and (b) input-output measurement using the interaction of the coupled cavity array with an soliton and a transmission line. (c) Sketch of the circuit proposal; curved lines represent superconducting resonators interrupted by a JJ (squares with cross and a capacitor.)}
\label{fig:in-out}
\end{figure}
%

In this work, we will discuss the existence of stationary solitons
within the DD-BH.  
Stationary solitons have an important advantage over exact
solutions of conservative equations.  Stationary solutions, if stable,
are obtained via the dissipative dynamics no matter of the initial
state (belonging to the basins of attraction).  Therefore their
preparation is easier and more robust.

Along this work, we first 
argue that for the DD-BH model quantum solitons
have no anti-continuous limit, i.e,  the uncoupled lattice system has a
{\it unique} stationary solution 
\cite{Drummond1980, LeBoite2013,LeBoite2014}.  This is important, since the single site DD-DNLS for the same parameter regime can have different fixed points in their irreversible dynamics.  This
is a qualitative difference and should alert to take some care in the  utilization of the
DD-DNLS for finding solitons.  To fix this problem, we include
quantum fluctuations up to second order. In doing so, we  recover the
uniqueness in the stationary single-site solution.  Moreover, solitons
exists within this (we  call it) \emph{semiclassical} approximation.  
We discuss their stability and range of existence. We also describe
other types of phases appearing when the soliton solution is unstable
or absent. We complement our study with a Gutzwiller-\emph{Ansatz}
showing that localization is more persistent within the Gutzwiller in
the range where solitons exist within the semiclassical limit. Finally, we
discuss a physical realization for the DD-BH based on a  circuit-QED
architecture \cite{Leib2012}. 
The  physical support for our model is complemented with a proposal
for a measurement scheme based on an input-output theory, to 
access the different phases using (already demonstrated)
experimental capabilities.

The rest of the paper is organized as follows.
The next section \ref{model} presents the model including dissipation and
gain.  Besides, we briefly describe the different theoretical
approaches used:  a second order (in the quantum fluctuations)
expansion (SOE)  and the Gutzwiller-Ansatz.  We finish the section with a possible implementation in circuit QED architectures. In Sect. \ref{results} we show our
numerical results in both approximation schemes and we compare them
against the \emph{classical} DD-DNLS. We present in Sect. \ref{inoutsec}
the input-output formalism for measuring the different phases and
conclude with some discussion in. \ref{conc}.   \\


\section{Model and its (approximate) solutions}\label{model}

The Bose Hubbard (BH) model  with driving reads ($\hbar = 1$) :
\begin{equation}
\label{H}
H=\sum_{n} \delta \omega a_n^\dagger a_n +\frac{U}{2} a_n^{\dagger 2}
a_n^2-J (a_{n+1}^\dagger a_{n}+ {\rm  h.c.})+ A (a_n^\dagger + a_n)
\; .
\end{equation}
It marks a minimal model for interacting bosons in a lattice.  
The model (in the rotating frame of the drive) is characterized via $\delta \omega$  (the detuning of the
bare resonator frequency $\omega_0$  from the pump frequency
$\omega_d$, $\delta \omega = | \omega_0 - \omega_d |$), $A$ (the driving amplitude for this coherent external driving),   $U$  (the onsite repulsion) and 
$J$ (the strength of the hopping among  sites).  
\emph {Phenomenologically},  single-particle-losses can be casted in a
Gorini-Kossakowski-Sudarshan-Lindblad master equation \cite{Rivas2011}
\begin{equation}
\label{master}
d_t \varrho = - i [H,\varrho] +\gamma\sum_n   a_n \varrho a_n^\dagger - \frac{1}{2}\left\{a_n^\dagger a_n,\varrho\right\} 
\end{equation}
with $\gamma^{-1}$ the time scale for the losses and $\{ A, B \} = A B
+ BA$ the anticonmutator.
A pictorial and
physical realization based on circuit QED is shown in fig. \ref{fig:in-out}. 
Without loss and driving, the BH  is a cornerstone in many body physics.  The generalized BH, Eqs. \eqref{H} and \eqref{master} mark, then,  a paradigmatic model for the study of  collective phenomena with driving and dissipation.
The dynamical equations for the averages $\langle a_n^\dagger ... a_m
\rangle \equiv {\rm Tr} (a_n^\dagger ... a_m \varrho )$  are given by,
\begin{widetext}
\begin{subequations}
\label{2ndo}
\begin{align}
\label{2ndo-1}
i  \frac{{\rm d} \langle a_n \rangle}{{\rm d}t}  = \; & ( \delta\omega - i \frac{\gamma}{2} )\langle a_n\rangle+
 A -J ( \langle a_{n+1}\rangle+\langle a_{n-1}\rangle ) 
\\ \nonumber
& +U ( 2\langle a^\dagger_n a_n\rangle\langle a_n\rangle+\langle a_n^2\rangle\langle a_n^\dagger\rangle -2\langle a_n\rangle^2\langle a_n^\dagger\rangle ) 
\\ 
\label{2ndo-2}
i\frac{{\rm d} \langle a_l^\dagger a_n \rangle}{{\rm d}t}  =& 
-i\gamma \langle a_{l}^\dagger a_n \rangle  + A ( \langle a^\dagger_l\rangle-\langle a_n\rangle ) 
 + J( \langle a_{l-1}^\dagger a_n \rangle+\langle a_{l+1}^\dagger a_n \rangle -\langle a_{l}^\dagger a_{n-1} \rangle-\langle a_{l}^\dagger a_{n+1} \rangle ) 
\\ \nonumber
& 
+ U ( \langle a_l^\dagger a_n^\dagger a_n a_n \rangle-\langle a_l^\dagger a_l^\dagger a_l a_n \rangle )
\\
\label{2ndo-3}
i  \frac{{\rm d} \langle a_l a_n \rangle}{{\rm d}t} = 
\; & (2 \delta\omega - i \gamma )  \langle a_l a_n \rangle
+A ( \langle a_l\rangle+\langle a_n\rangle )  
-J ( \langle a_{l-1} a_n \rangle+\langle a_{l+1} a_n \rangle+\langle a_{l} a_{n-1} \rangle+\langle a_{l} a_{n+1} \rangle ) 
\\ \nonumber
& + U ( \langle a_l^\dagger a_l a_l a_n  \rangle +\langle a_n^\dagger a_n a_n  a_l\rangle
+\delta_{n,l}\langle a_n a_n\rangle ) 
\\ \nonumber
\\ \nonumber
\\ \nonumber
\vdots
\end{align}
\end{subequations}
\end{widetext}
The dots above indicate that, due to the interaction term $U a_n^{\dagger 2} a_n^2$, an endless hierarchy of
equations for the $n$-point
correlators $\langle a_n^\dagger ... a_m
\rangle$  is obtained.  Therefore, the set needs to be cut at some order.

\subsection{Zeroth order: The DNLS equation}
 
The simplest approximation is the so-called \emph{classical} limit,
consisting in replacing operators by their averages:  $\langle
a_n^{\dagger 2}a_n\rangle \to |\varphi_n|^2\varphi_n$, with $\varphi_n=\langle a_n\rangle$.
The approximation can be understood as the zeroth order cumulant
expansion in the quantum fluctuations.  In doing so, the equation for
the first moments (\ref{2ndo-1}) forms already a closed set.
The resulting equations are the celebrated 
DNLS equations, in this case, with driving and dissipation: 
\begin{equation}
\label{dnls}
id_t \varphi_n=\delta\omega \varphi_n+U|\varphi_n|^2\varphi_n-J(\varphi_{n+1}+\varphi_{n-1})+A-\frac{i\gamma}{2}\varphi_n \, .
\end{equation}
For this set of equations \cite{Peschel2004, Egorov2013},  apart from dark soliton and kinks, there also exist bright solitons for a defocusing nonlinearity $U=-1$, on which we will focus along this work.  The main question that we tackle is, if the solutions found in the DNLS  survive the inclusion of quantum fluctuations.


\subsection{Second order expansion }

Let us consider now 
equations \eqref{2ndo} up to 
second order of correlations. 
In doing so, we rewrite
\begin{equation}
\label{ansatz}
\hat{a}_n= \hat{a}_n-\langle a_n\rangle+\langle a_n\rangle=:\delta
\hat{a}_n+\langle a_n \rangle \, .
\end{equation}
Neglecting  terms with $\mathcal{O}(\delta \hat{a}^3)\simeq0$, any
$n$-correlator can be written in terms of $2$-point correlators:
\begin{eqnarray}
\label{BA}
\langle A_1 A_2 A_3 A_4 \rangle
&=
\sum_{\substack{
 j < k 
\\
 l< m} 
  }
\langle A_j A_k \rangle
\langle A_l \rangle \langle A_m \rangle
\\ \nonumber 
& 
-
5
\langle A_1 \rangle \langle  A_{2} \rangle
\langle A_3 \rangle \langle A_4\rangle
\; .
\end{eqnarray}
Above $j \neq k \neq l \neq m$ and $A_j$ can be any annihilation $a_n$
(creation $a_n^\dagger$) operator.
Consequently,  equations \eqref{2ndo-1}, \eqref{2ndo-2} and \eqref{2ndo-3} form a
closed set which stands for a second order expansion (SOE). Its
numerical solution provides us the results in the subsection
\ref{ssec-soe}. 
It is worth emphasizing that, instead of the SOE,
Gaussian expansions as the Hartree-Fock-Bogoliubov (HFB) or higher
order terms might also be considered  \cite{Tikhonenkov2007,Witthaut2011, Quijandria2014}.
However, in the parameter regime explored, our SOE  approaches better
the exact result for the single site case
\cite{Drummond1980}.

\subsection{Gutzwiller \emph{Ansatz}}
\begin{figure}[t]
\centerline{\includegraphics[width=\columnwidth]{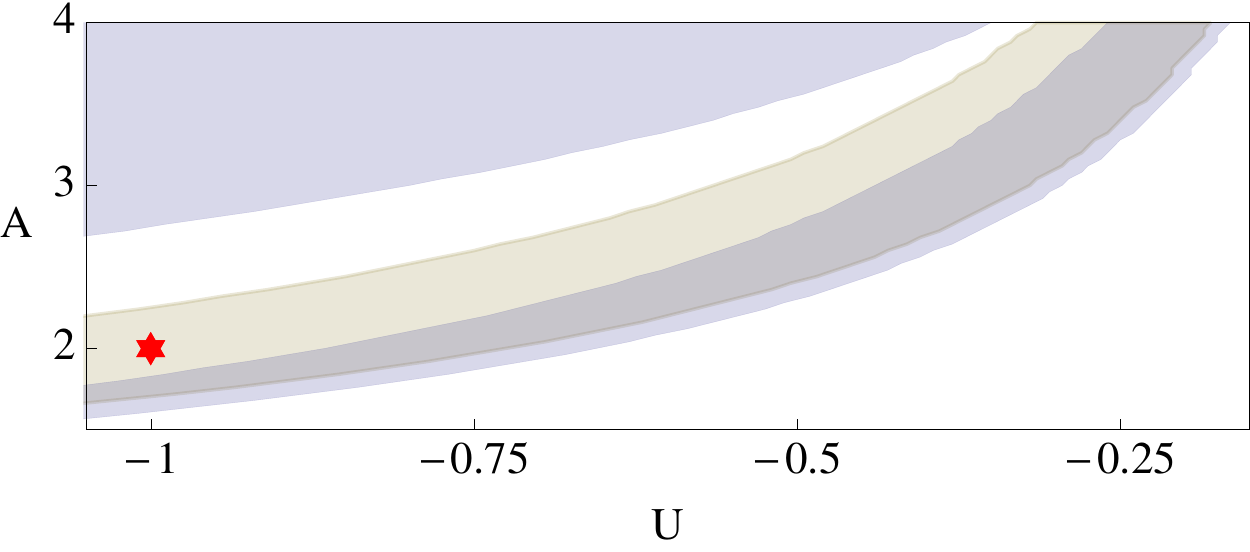}}
\caption{(Color online) Number of single-site solutions vs. $U$ and $A$ for the DNLS and SOE approximation. As explained in the text, forbidden parameter regimes have more than one solution and are shown for DNLS (light gray) and SOE (dark gray). The red star denotes the parameters used along the letter.  }\label{fig2}
\end{figure}
  
To consider higher on-site correlations, we will compare the
results of SOE  with the time evolution of a density
matrix using a Gutzwiller \emph{Ansatz} \cite{Krutitsky2010,
  Quijandria2014} (which assumes a factorized form for the density matrix): 
\begin{equation}
\label{gtzwrho}
\varrho=\prod\limits_{n=1}^N   \bigotimes \varrho_n \, ,
\end{equation}
 with site-dependent density matrices  $\rho_n$.
Using that ${\rm tr} ( \varrho_n)=1$ (${\rm tr} (\partial_t
\varrho_n) = 0$), we obtain the quantum nonlinear master equation set:
\begin{align}
\partial_t \varrho_n  =- i\Bigg[ &
\delta\omega \,  a_{n}^\dagger a_{n} +A(a_n^\dagger+a_n)
+ \frac{U}{2} a_{n}^{\dagger 2}
  a_{n}^2
\nonumber \\ 
&-J \Bigg( \big (\langle a_{n+1}^\dagger\rangle+\langle
    a_{n-1}^\dagger\rangle )a_{n}+{\rm h.c.}\Bigg ) , \rho_n\Bigg ]
\nonumber \\
& \hspace{-4.ex} +\gamma
a_{n} \rho_n a_{n}^\dagger -\frac{1}{2}\left\{a_{n}^\dagger
  a_{n},\rho_n\right\}
\label{gtzw}
\end{align}

\subsection{Circuit-QED implementation}

Though several systems may be modeled by means of a BH model with losses and
external driving as in Eqs. \eqref{H} and \eqref{master}, we fix our
attention on circuit QED architectures \cite{Jin2013}.  
The latter seem to be an ideal platform  to study many body physics \cite{Houck2012}.  
In this subsection we argue that the fundamental blocks for simulating
\eqref{H} have been already experimentally demonstrated.

The first ingredient is having nonlinear resonators.  For that,  we
think about recent experiments where coplanar waveguide resonators are interrupted by
a Josephson Junction (JJ)  [Cf. Fig. \ref{fig:in-out}c]. 
The JJ provides the nonlinearity through the term $E_J \cos (2 \pi
/\Phi_0 \delta \phi)$ in the effective action. Here $E_J$ is the Josephson energy, $\Phi_0$ the
flux quanta and $\delta \phi$ the \emph{jump} in the flux at both
sides of the junction \cite{Zueco2012, Bourassa2012}. 
In 
Ref. \onlinecite{Ong2011}, the authors measured nonlinear resonators that can be
modeled within the Hamiltonian (after expansion of the cosine):
\begin{equation}
H = \omega_0 a^\dagger a+ \frac{U}{2} (a^\dagger )^2 a^2 + \frac{U^\prime}{2}
(a^\dagger)^3 a^3
+ \dots
\end{equation}
with $\omega_0 \cong 6 {\rm GHz}$ and $U \cong -700 \rm{KHz}$.
Therefore, by choosing pumps with  driving frequencies detuned from the
$\omega_0$ on the $\rm KHz - MHz$ regime different $U /\delta \omega$
in (\ref{H}) can be simulated.   Finally, higher order terms can be safely discarded,  $U^\prime / U \cong 10^{-3}$.
An intercavity coupling as in (\ref{H}):
\begin{equation}
H = J  ( a^\dagger_n a_{n+1} + {\rm h.c.} )
\end{equation}
has been already measured in a wide range of values for $J$, even
reaching values of $J/ \omega_0  \cong 0.2$ \cite{Haeberlein2013}. Moreover, a tunable coupling $J$ has been measured \cite{Baust2014}. Highly
reproducibility in the resonator bare frequencies,  a necessary
ingredient for building many body arrays, 
has been also achieved \cite{Underwood2012}.

Finally,  a measurement scheme is mandatory.  Here,  we rely on
the field tomography techniques developed in the circuit QED community \cite{ Menzel2010,
  Bozyigit2010, Eichler2011}.  
As explained in section \ref{inoutsec}, measuring field-field
correlators is sufficient for accessing the different phases of
(\ref{H}), including the solitonic solutions.
A possible architecture is depicted in figure \ref{fig:in-out} c).
Inspired on Ref. \onlinecite{Leib2012} we envision a one-dimensional
array of nonlinear cavities:  superconducting resonators interrupted
by a JJ.  The design is such, that  the coupling can be
tuned by  locally approaching the resonators
\cite{Haeberlein2013}.  The measurement can be accomplished by an
auxiliar transmission line that couples the array and  where an input
field is impinged and the ouput is measured as we will explain in sec.
\ref{inoutsec}.
Therefore, the simulation and measurement may be possible within
the technological state of the art.


\section{Results}\label{results}

We summarize here our numerical findings. We first review the \emph{classical}
DNLS limit.  Then, we report on our quantum results, both for the SOE
and Gutzwiller-Ansatz.

\begin{figure*}[t]
\centerline{\includegraphics[width=2\columnwidth]{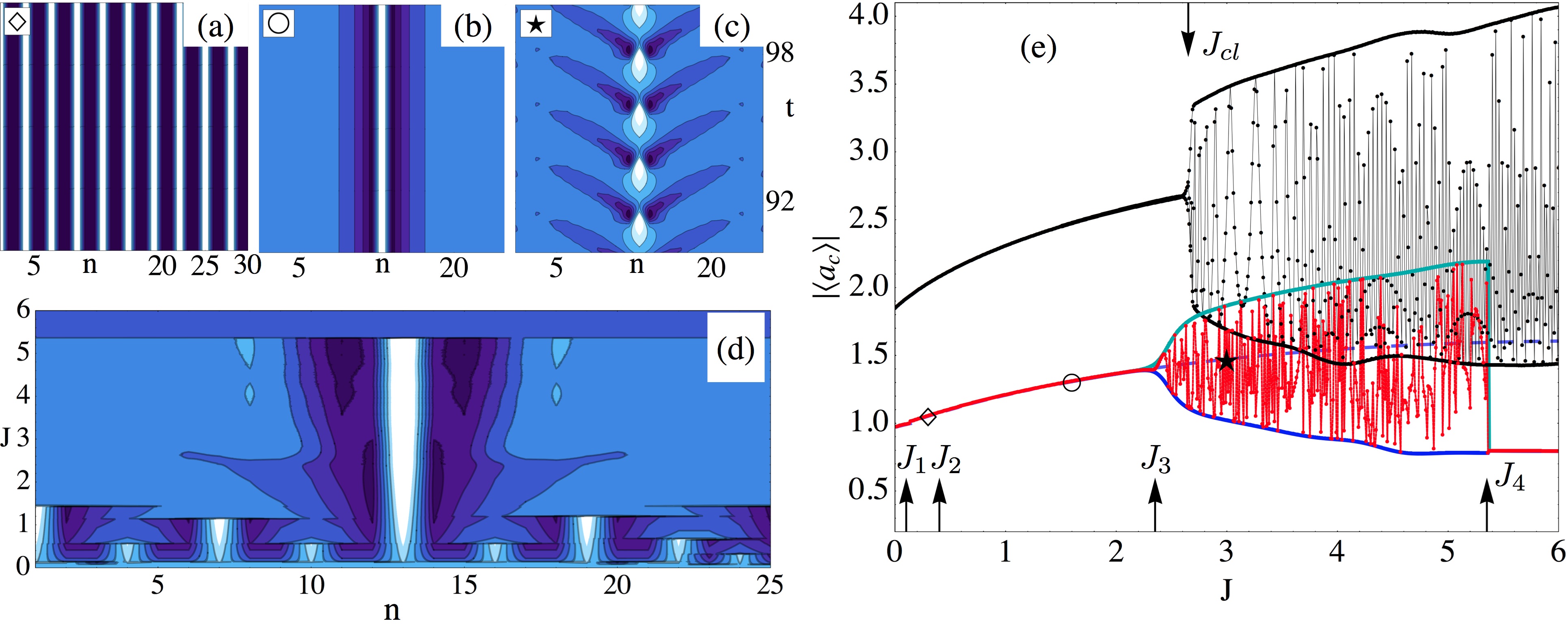}}
\caption{(Color online) Lont-time dynamics of $\langle a_n\rangle$ vs. $n$ and $t$ under SOE, showing ripples for $J=0.3$ (a), a stable soliton for $J=1.6$ (b) and the oscillatory localized mode for $J=3$ (c). In (d) SOE steady state of \eqref{2ndo} vs. sites $n$ and $J$, in the oscillatory regime $0.5\left(|\langle a_n(t_{min})\rangle|+|\langle a_n(t_{max})\rangle|\right)$.   (e) Amplitude of the center site for DNLS (black), SOE Max(green), SOE Min (blue) and SOE fixed integration time (red) and dashed blue line: unstable localized stationary mode.  }\label{fig3}
\end{figure*}

\subsection{DNLS: solitons with anti-continuous limit}

When considering the DNLS one common procedure for finding
localized modes is as follows.
Imagine that two stable solutions exist for the single site case, say
$\varphi$ and $\varphi^\prime$.
Then, at zero hopping ($J=0$), the solution $\varphi_m =
\varphi^\prime$, $\varphi_n= \varphi$ $\forall n \neq m$ is  a
 stable localized solution.  This example corresponds to a soliton
with a localized amplitude at one site ($m$).
This (trivial) soliton can be used as starting point to find solutions
by turning on the hopping, $J\neq 0$.  
In the nonlinear jargon, the zero hopping case is named as anti-continuous limit and \cite{MacKay1994, Flach2008, Kevrekidis} a variety of numerical continuation techniques
from zero to non-zero hopping have been used as, for example, the
Newton-Raphson method. This procedure is not restricted to conservative settings, but also works very well in the driven dissipative
classical models mentioned before. 
So was used in \cite{Peschel2004, Egorov2013} to
characterize the localized modes for the DD-DNLS.
The results of this numerical continuation for the DD-DNLS \eqref{dnls} will be compared to SOE in the following.

\subsection{SOE: Localization without anti-continous limit}\label{ssec-soe}

In the quantum regime, the continuation from the zero hopping (anti-continuous limit) can not be used.
The reason is, that the single site of \eqref{H} and \eqref{master}
has a unique solution \cite{Drummond1980}.  
Therefore, \emph {if quantum solitons exists they do not have an anti-continuous limit.}
Without the possibility of finding solitons by continuation, an
educated guess is to try the search  within the
parameter regime where they exist in the classical DNLS limit
\cite{Peschel2004, Egorov2013}.
As anticipated, through this work we use the approximate treatment
SOE.  Therefore, for consistency, we must check
that SOE has a unique solution in the single site case.   In figure
\ref{fig2} we delimit this consistency region. Dark-gray areas
denote multivalued SOE solutions of \eqref{2ndo} and therefore
forbidden regions.  For comparison we
also draw (in light-gray) the forbidden space for the DD-DNLS.  The
red star marks a point with a unique solution for the SOE but with
proven solutions in the DNLS.  It seems a good point to start with. 
The concrete parameters are  $U=-1$ $A=\gamma=2$ and  -for a better
comparison with the DNLS \cite{Peschel2004, Egorov2013}- detuning of
$\delta\omega=3+2J$. The only free parameter remaining is the coupling
$J$.   Those parameters are used along the text.

In general, steady-state solutions can be obtained by simply
integrating the dynamics for \eqref{2ndo} up to sufficiently long
times such that a stationary dynamics is reached\footnote{We check that the condition 
\unexpanded{ $
\sum_n| \langle
 a_n(t)\rangle -\langle a_n(t+\Delta t)\rangle|\leq 10^{-7} 
$}
 is
  fulfilled for the steady-state. For time-periodic solutions we use
  the corresponding condition for the comparison of two consecutive
  distributions with maximal center site amplitude.}.
By construction,
only stable solutions are found. 
Besides, 
there is no necessity of fine-tuned state preparation.
Finally, this method  provides not only 
 steady-state solutions,  but also time-periodic modes. 
Another
possibility, which also finds unstable modes, is using the corresponding
algebraic set of equations for $d_t \langle \dots \rangle=0$ with
e.g. a Newton-Raphson scheme. Unfortunately, this method has no
guarantee to converge and could be used successfully only for specific
parameters (see the unstable soliton mentioned below). We use the long-time dynamics for all stable modes presented here.

Examples of different  solutions  assuming periodic boundary
conditions are plotted in fig. \ref{fig3}.
In figure \ref{fig3}a  the dynamics for a stationary ripple mode is depicted.  In
Fig. \ref{fig3}b) and c) examples for the time evolution of stationary and periodic localized modes are plotted respectively. 
To visualize the physical mechanism yielding these solutions we choose
to plot the mean amplitudes $0.5\left(|\langle
  a_n(t_{min})\rangle|+|\langle a_n(t_{max})\rangle|\right)$ vs. sites
$n$ and $J$  in Fig. \ref{fig3}d. This averaging is recommendable to
better illustrate the  localized character of the oscillatory mode, in
all other regions of steady-state modes it does not change the
picture. 
The figure shows that, 
for vanishing and small $J$, the homogeneous mode is the only stable
solution. It becomes unstable at $J_1\simeq 0.1$, a
symmetry-breaking bifurcation not present in the DNLS limit. For
small, but finite $J>0.1$ the ripple modes with one site having a
higher amplitude than its two neighbors dominate the dynamics. For the
spatial periodicity of these modes there is a certain dependency on
the number of sites, as can be seen in fig. \ref{fig3}(d) with $n=25$
leading to a defect in the right bottom corner, whereas for $n=30$ in
fig. \ref{fig3} (b) no such defect can be found. As the coupling
increases, repeated bifurcations into modes with different periodicity can be observed as
the extension of the maxima grows and less peaks can be accommodated
within the lattice. Finally, this leads to only one central localized
mode in fig. \ref{fig3}(d) at $J_2$ appearing dynamically as the steady state. 
Increasing $J$, the stationary soliton starts to oscillate at $J_3$ and finally relaxes to the homogeneous mode for  $J_4\gtrsim 5.38$. Thus, the SOE does exhibits a train of symmetry breaking bifurcations towards more and more localization, a behavior not found in the DD-DNLS.

We have seen that the classical DNLS is a particular limit of the quantum model in which the quantum fluctuations are neglected. If the quantum corrections were negligible, both the SOE and the DNLS would produce similar results. As shown in Fig.  \ref{fig3}(e) this is not the case. There we plot  the dependence of the center site amplitude $|\langle a_c(t_{evol})\rangle|$ on $J$ after the dynamics settled into a steady-state or periodic state of SOE (\eqref{2ndo} (red) within the evolution time $t_{evol}=100$.  Please note, that all possible phases are shown in \ref{fig3}(e) and  the value of $|\langle a_c\rangle|$ does not necessarily indicate that there is a difference to neighbouring sites. The arrows point to the bifurcation points $J_i$. When the dynamics is determined by the periodic mode we also show the maximum and minimum of  $|\langle a_c(t)\rangle|$ in green and blue. The algebraic stationary and unstable localized mode is shown with a blue dashed line. For comparison we show the results of the classical limit \eqref{dnls} in black, also exhibiting a periodic mode, but for higher $J$. The classical amplitude is nearly twice as high as the SOE value, but the main difference is in the classes of solutions found. Whereas the soliton mode is stable in the classical DNLS limit from $J=0$ up to the appearance of the periodic solution at $J_{cl}$, for the SOE limit there is no anti-continuous limit. At $J_1$ the ripples appear and persist for $J_1<J<J_2$. The bifurcation into the periodic solution is located at $J_3<J_{cl}$; as well as the high-coupling homogenous mode at $J_4$. The symbols in the SOE families denote the examples shown in \ref{fig3}(a)-(c).

\subsection{Gutzwiller \emph{Ansatz}}

\begin{figure}[t]
\centerline{\includegraphics[width=\columnwidth]{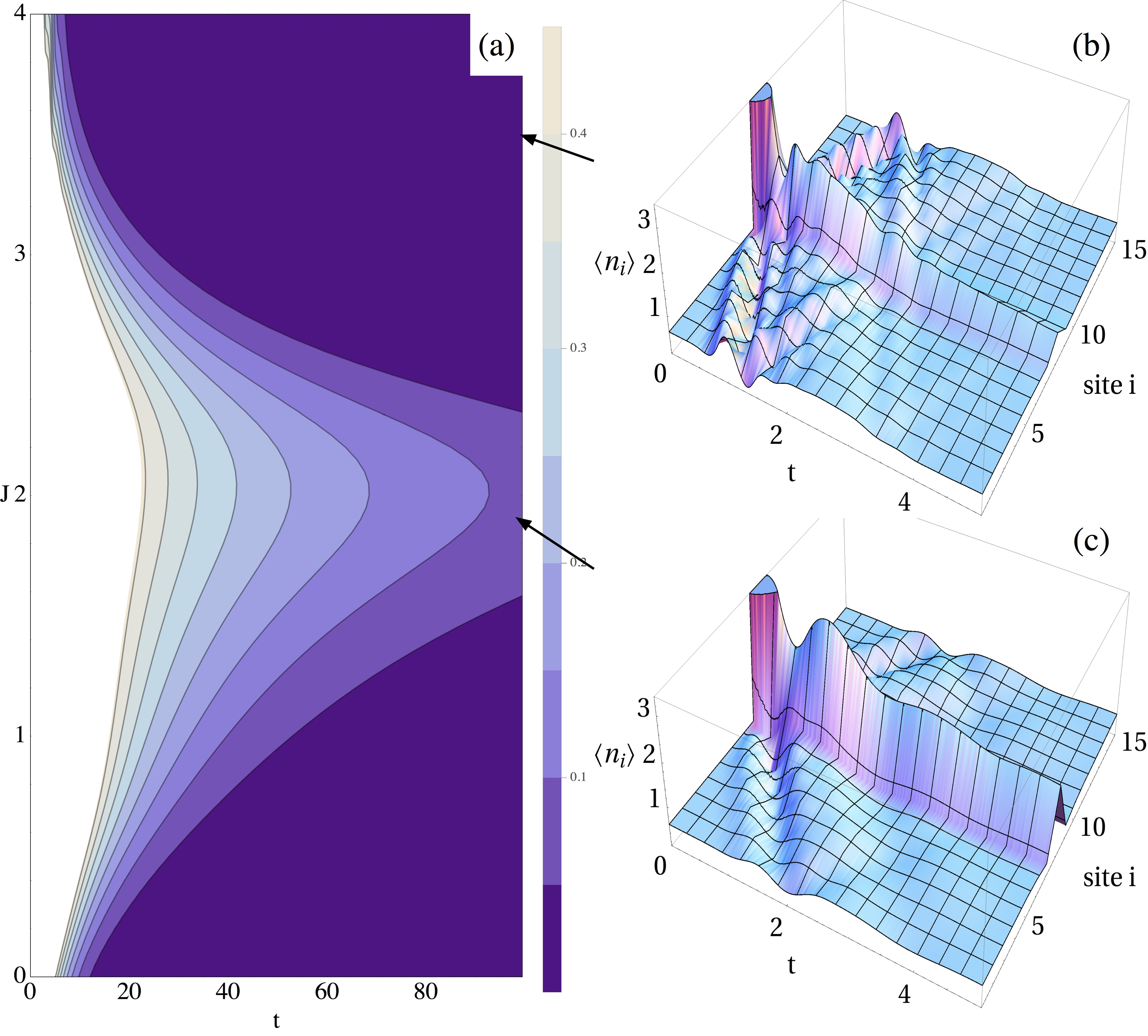}}
\caption{(Color online) Gutzwiller-Ansatz \eqref{gtzw}:  (a) $\langle n_c(t)\rangle-\langle n_1(t)\rangle$. Left: ${\langle n_i(t)\rangle}$ for $J=2$ (b) and $J=3.5$ (c)}\label{fig4}
\end{figure}

We complement the SOE with the dynamics within the Gutzwiller {\it
  Ansatz} \eqref{gtzwrho}. 
As initial conditions, we use the homogeneous steady state for the
corresponding value of  $J$  at all sites but the center, where we
assume a coherent state with higher $\langle n_c(t=0)\rangle\simeq
8.8$. The Fock-space per lattice site $\rho_i$ is truncated to a
maximum of $15$ excitations in a lattice of $15$ sites. Within this
parameter space, we were able to find a region at $J\simeq2$, where the initially localized distribution survives much longer. Even though we can compare the SOE and the Gutzwiller {\it  Ansatz} only qualitatively, this corresponds to the regime of stable soliton modes in the SOE limit. 
In fig. \ref{fig4}(a) we show the time evolution of \mbox{$\Delta n_{c,1}(t)= \langle n_c(t)\rangle-\langle n_1(t)\rangle$} vs. $t$ and $J$, which gives indications about the survival time of localization.  Whereas for small $J$ the value of $\Delta n_{c,1}(t)$ decays very fast, the relaxation to the homogeneous state for $J\simeq 2$ is much slower. The values of $\langle n_i\rangle$ in the whole array for $J=3.5$ and $J=2$ are plotted in fig. \ref{fig4} (b) \& (c), respectively. Fig. \ref{fig4}(c) shows, that the initial value decays abruptly to $n_c\simeq 3$, from that point on the decay becomes very slow. This indicate the existence of a weakly unstable localized mode. For the example shown in fig. \ref{fig4}(b), the initial decay is equally fast, furthermore some oscillations can be observed, which is a reminiscence to the existence of the periodic mode in the SOE limit for these parameter values. Additionally, hints of these oscillations can be seen in \mbox{$\Delta n_{c,1}$ } for $J\simeq 3.5$ . We could not find any indications of an ripples modes, since it presents higher-order inter-site correlations neglected within the Gutzwiller-Ansatz, as we will show in the following section.

\section{In-Out mechanism}\label{inoutsec}

It still remains the question of how to extract the information stored
in our discrete array of cavities. In order to do this, we will follow
the {\it input}-{\it output} formalism \cite{Gardiner1985}. The basic
idea here is to make our dissipative system interact with the electromagnetic
field (EM), in the form of a transmission line (TL) [Cf. figure
\ref{fig:in-out} b) and c)].
The EM field can be decomposed into two
contributions: the {\it input}, or the radiation impinging onto the
system, and the {\it output}, the sum of a reflected plus a radiated
component [See Fig.\ref{fig:in-out}(b)]. The latter is  determined by the
system and its interaction with the TL.
Therefore,  measuring the output (and comparing it with the input) we can 
infer information on the system dynamics. 

We will briefly sketch the main steps to derive the \emph{input}-\emph{output} relations. Details on the calculations can be found in Appendix \ref{app:a} (see also the Supplementary Material
of Ref. \cite{Quijandria2013}). 
The system we want to acquire information from is an \emph{open}
system described by the Hamiltonian (\ref{H}) and the master equation
(\ref{master}).  Let us denote this open system: the BH model +
dissipation and driving as $H_{open}$. 
The open system $H_{open}$ is coupled to a transmission line as
depicted in figure \ref{fig:in-out} c).  The   total Hamiltonian,
including the TL and the interaction of our system with it, reads $H =
H_{\rm open} + H_{\rm TL} + H_{\rm int} $. 
The TL interacts directly with the cavities and it can be viewed as a
one dimensional  EM field. In second quantization,
the  EM field can be described as a collection of harmonic oscillators. As it is depicted in fig. \ref{fig:in-out} b), we should consider carefully the direction of propagation of excitations in the TL. For this task we will introduce the EM field operators: $l(p)$ ($l^{\dagger}(p)$) which destroys (creates) a photon with momentum $p$ propagating to the left and $r(p)$ ($r^{\dagger}(p)$) which destroys (creates) a photon with momentum $p$ propagating to the right. In terms of these and for a linear dispersion relation ($\omega_p = v \vert p \vert $), the Hamiltonian of the EM field in the TL reads
\begin{equation}
H_{\rm TL} = v \left( \int_{0}^{\infty} {\rm d}p\, v p \, r^{\dagger}(p) r(p) - \int_{-\infty}^{0} {\rm d}p\, v p \, l^{\dagger}(p) l(p) \right)
\end{equation}
with $v$ the speed of light in the TL. Finally, the interaction $H_{\rm int}$ considers
the most general type of coupling in a solid-state device. It 
consists of an inductive part (flux interaction) and a capacitive
contribution (charge interaction). The interactions are weak and
point-like, happening at the position of every cavity, yielding 
\begin{align}
H_{\rm int} & \sim  \frac{ig}{\sqrt{2\pi N}} 
\sum_k  \left(  \int_{+k} \frac{{\rm d}p}{\sqrt{\omega_p}} \, N  \left( r^{\dagger}(p) a_k(t)  - {\rm h.c.} \right) \right. \nonumber\\
 & +\left.  \int_{-k} \frac{ {\rm d}p}{\sqrt{\omega_p}} \, N  \left( l^{\dagger}(p) a_k(t)  - {\rm h.c.} \right) \right)
\end{align}
where we have introduced a plane wave expansion for the cavity operators
\begin{equation}
a_n(t) = \frac{1}{\sqrt{N}} \sum_k {\rm e}^{ik x_n /d} a_k(t)
\end{equation}
$d$ being the lattice spacing of our cavity array with $N$ sites. We are now able to solve the Heisenberg equations for the left and right operators. The idea is to relate the contributions of all momenta before the interaction (from an initial time $t_0 \rightarrow -\infty$) and after the interaction (up to a time $t_1 \rightarrow \infty$). As it is shown in appendix \ref{app:a}, for every cavity momentum $k$ we only have significant contributions from those momenta $p$ in a narrow region around the former. For the right operators, we call this momentum interval $+k$ and for the left ones $-k$ [cf. eq. \eqref{A8b}]. Taking this into account, we introduce the (right) \emph{input} operator
\begin{eqnarray}
r^{+k}_{\rm in}(t) = \frac{1}{\sqrt{2\pi}} \int_{+k} {\rm d}p \, {\rm e}^{-i vp(t-t_0)} r_0(p)
\end{eqnarray}
for times $t>t_0$ ($r_0$ denotes the $r$ operator at $t=t_0$) and the (right) \emph{output} operator
\begin{eqnarray}
r^{+k}_{\rm out}(t) = \frac{1}{\sqrt{2\pi}} \int_{+k} {\rm d}p \, {\rm e}^{-i vp(t-t_1)} r_1(p)
\end{eqnarray}
for times $t<t_1$ ($r_1$ denotes the $r$ operator at $t=t_1$). And similarly for the $l$ operators. Thus, the Heisenberg equations lead to the following \emph{input}-\emph{output} relations
\begin{eqnarray}\label{rout}
r^{+k}_{\rm out}(t) = r^{+k}_{\rm in}(t) + \frac{1}{d} \sqrt{\frac{\Gamma v}{N k}} {\rm e}^{-ikvt/d} a_k
\end{eqnarray}
\begin{eqnarray}\label{lout}
l^{-k}_{\rm out}(t) = l^{-k}_{\rm in}(t) + \frac{1}{d} \sqrt{\frac{\Gamma v}{N k}} {\rm e}^{+ikvt/d} a_k
\end{eqnarray}
with $\Gamma$ a constant characterizing the strength of the
TL-nonlinear cavity coupling, and $a_k$ defined as
\begin{equation}
a_k = \frac{1}{\sqrt{2\pi}} \int {\rm d}t' \, {\rm e}^{ikvt'/d} a_k(t')
\end{equation}
In the presence of the TL the equations of motion for the $a_k(t)$
operators differ from those obtained from (\ref{master}). The TL plays
now the role of a second environment for the system described by
(\ref{H}). However, under the same approximations which led to
(\ref{master}) (Markov approximation), we immediately see that the
role of the TL is to renormalize the decay rates $\gamma$. Namely, to
add a contribution proportional to $g^2$ to them. In addition, the
input field will renormalize the driving field amplitude (with strength of order $g$). 

\begin{figure}[t]
\centerline{\includegraphics[width=\columnwidth]{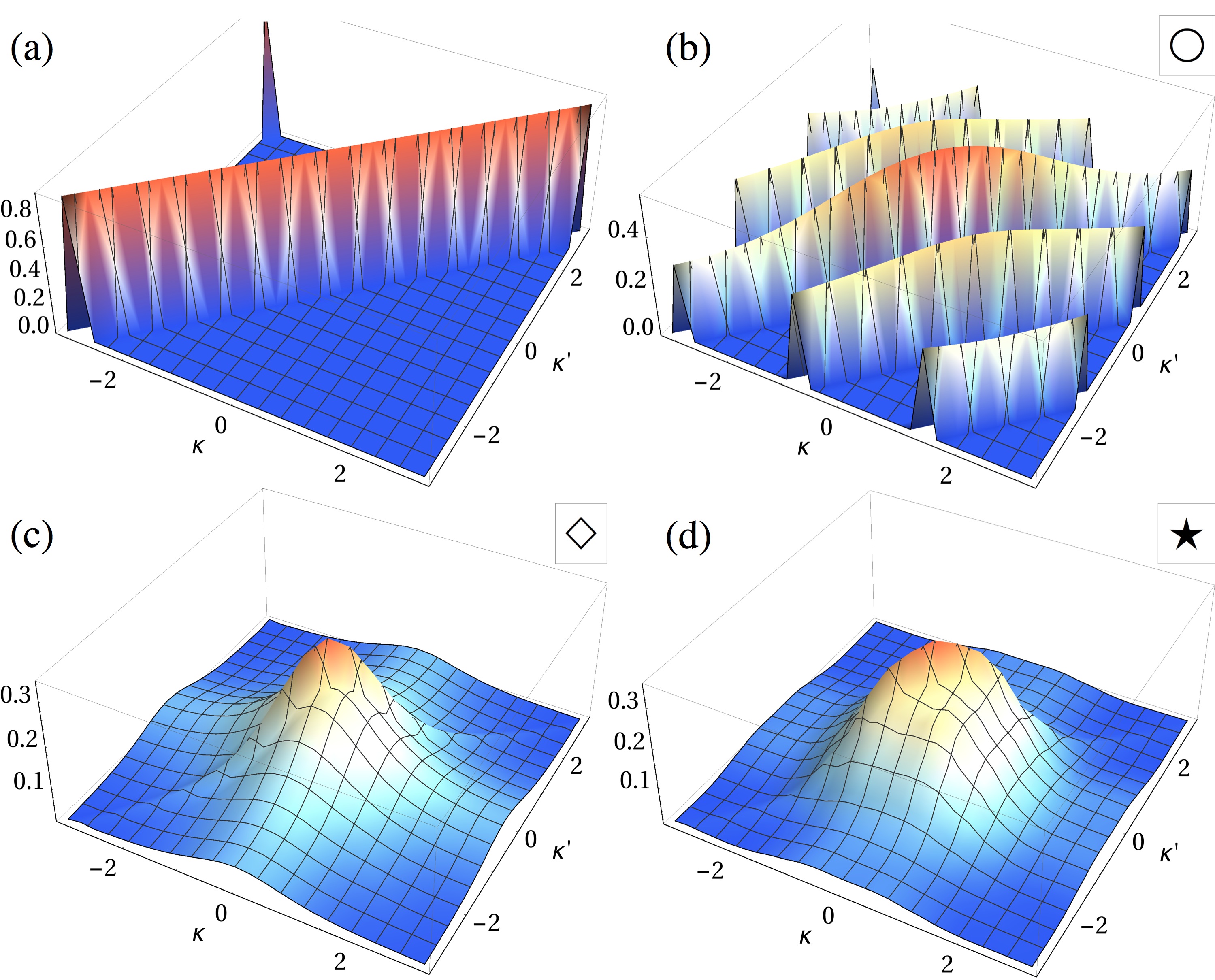}}
\caption{(Color online) Connected correlator $f\left(a_{k},a_{k'}^\dagger\right)$ \eqref{2pfun} for $J=0, (0.3, 2,3)$ (a)-(d). For (b)-(d), the symbols correspond to the parameter values and solutions shown in fig. \ref{fig3}(a)-(c).}\label{fig5}
\end{figure}

Within relations (\ref{rout}) and (\ref{lout}) we can map output field-field
correlations to cavity  modes correlations.
For example, using  relation (\ref{rout})  $\langle r_{\rm
  out}^{+k \dagger} (t_1) r_{\rm out}^{+k} (t_2) \rangle$ gives us
information on the system two-point correlator $\langle a_k^\dagger
a_{k^\prime} \rangle$ provided that we only impinge a signal from the left
\begin{equation}
\langle r_{\rm out}^{+k \dagger } (t_1) r_{\rm out} (t_2) \rangle =  \frac{1}{d^2}\left( \frac{\Gamma v}{N \sqrt{ k k'}} \right) {\rm e}^{iv(kt_1 - k't_2)/d} \langle a_k^{\dagger} a_{k'} \rangle 
\end{equation}
Similar relations hold for other two-point correlations. 



Therefore, two time correlations in the output field map to the system two point
correlations in momentum space.  The first are experimentally
accessible, the latter as we will show now provide sufficient
information for distinguishing the different phases described throughout
this work.  We introduce the connected correlator,
\begin{equation}
\label{2pfun}
f(a,b)=\langle a b\rangle-\langle a \rangle \langle b\rangle
\, .
\end{equation}
It is worth emphasizing that the above is identically zero in the
classical limit. 
In fig. \ref{fig5} we plot 
$f\left(a_{k},a_{k'}^\dagger\right)$
for different stationary SOE solutions.
The homogeneous solution ($J=0$) is shown \ref{fig5} a).  The ripples
($J=0.3$) are drawn in \ref{fig5} b) with the same parameters as in
fig. \ref{fig3} a).  Localized solutions, static and oscillatory are
given in  \ref{fig5} c) and d) respectively.  Their real space
counterparts are given in \ref{fig3} b) and c).  
The momentum space for each phase   are clearly distinguishable from each
other and show, that the input-output mechanism  presents a perfect measurement scheme to prove the existence and stability of localized modes. \\

\section{Discussion}\label{conc}

The phases for the Bose-Hubbard model with gain and loss have been
investigated within a semi-classical approach.
It has been argued that, in the zero hopping case, the unique
solution is  the homogeneous one.  A translational-invariant
broken symmetry solution (ripples) appears when the hopping term reaches some
critical value. Increasing the hopping, the extension of the ripples
grows and their periodicity decreases in a second bifurcation. Eventually, the discrete periodicity disappears and only one
 maximum remains, the stable localized mode. Passing from static to periodic (in time), this  mode finally becomes unstable at higher $J$, transiting to a homogenous solution.

Those successive symmetry-breaking transitions (from homogenous, to
discrete periodic, to localized static and periodic modes, and back to homogenous) mark  novel
phases without counterpart in the Hamiltonian limit (zero dissipation,
zero gain) of the Bose-Hubbard where the well known Mott-superfluid
transition has been largely described.
Apart from the interest in finding novel matter phases in the many body phase diagram,  artificial systems with driving and dissipation present also a natural way of observing localization.

Our calculations rely on a semiclassical
approximation.  We have complemented them with a Gutzwiller Ansatz where
the on-site dynamics is expected to be more accurate but inter-site
correlations are poorly described. Nevertheless, the regions where
solitons exists in the semiclassical regime present long lived
localized solutions within the Gutzwiller-Ansatz.  Therefore, we expect that the semiclassical phases have some traces in the full,  not yet explored, quantum dynamics.

The richness of phases presented here may
be a motivation for future works considering the full quantum
aspects of the model.  In this line, our proposal within circuit QED
presents a quantum simulator for going beyond the theory presented here.

\section*{Acknowledgements}
We acknowledge support from the Spanish DGICYT under Projects No. FIS2012-33022 and No. FIS2011-25167 as well as by the 
Arag\'on (Grupo FENOL) and the EU Project PROMISCE. \\\\

\begin{appendix}

\section{\emph{Input}-\emph{output} theory}
\label{app:a}

Here we derive the \emph{input}-\emph{output} relations for the dissipative driven Bose Hubbard model coupled to a transmission line (TL). The total Hamiltonian reads:
\begin{equation}\label{A1}
H = H_{\rm open} + H_{\rm TL} + H_{\rm int}
\end{equation}
where $H_{\rm open}$ accounts for the open system: BH model + driving + environment. For simplicity, we will refer to the latter as our system. $H_{\rm TL}$ describes the EM field propagating through the TL. In momentum space it is given by
\begin{equation}\label{A2}
H_{TL} = v \left( \int_{0}^{\infty} {\rm d}p\, vp \, r^{\dagger}(p) r(p) - \int_{-\infty}^{0} {\rm d}p\, vp \, l^{\dagger}(p) l(p) \right)
\end{equation}
where we are assuming a linear dispersion relation $\omega_p = v \vert p \vert$. Our cQED proposal involves impinging a signal into the system and gathering information about it by means of the reflected and transmitted components of the former. For this task, it results helpful to decompose the EM field operators in: $l^{\dagger}(p)$ ($l(p)$) which creates (annihilates) an excitation with momentum $p$ propagating to the left with velocity $v$ (the speed of light in the TL) and similarly, $r^{\dagger}(p)$ ($r(p)$) which creates (annihilates) an excitation with momentum $p$ propagating to the right with velocity $v$. Finally, $H_{\rm int}$  is the interaction Hamiltonian 
\begin{equation}\label{A3}
H_{\rm int} = i \sum_n \int {\rm d}x\,g_n(x) \hat{A}(x) (a_n - a_n^{\dagger})
\end{equation}
Here we are considering a generic coupling between a system operator and the EM potential $\hat{A}(x)$
\begin{eqnarray}\label{A4}
\hat{A}(x) &=& v\left( \int_{-\infty}^{0} {\rm d}p \sqrt{\frac{1}{2\pi \omega_p}}  l(p) {\rm e}^{ip x} \right. \nonumber\\ &+& \left.   \int_{0}^{+\infty} {\rm d}p \sqrt{\frac{1}{2\pi \omega_p}}  r(p) {\rm e}^{ip x} +  {\rm h.c.}  \right)
\end{eqnarray}
In \cite{Quijandria2013} following the lumped circuit element description of a TL, we decomposed the interaction into capacitive and inductive  contributions. The latter are encoded in the coupling function $g_n(x)$. We now introduce a plane wave expansion for the cavity operators $a_n$
\begin{equation}\label{A5}
a_n(t) = \frac{1}{\sqrt{N}} \sum_k {\rm e}^{ik x_n /d} a_k(t)
\end{equation}
where $d$ is the lattice spacing of our cavity array with $N$ sites. Assuming a rotating wave approximation (RWA) regime we can rewrite (\ref{A3}) as
\begin{align}\label{A6}
H_{\rm int} &= \frac{iv}{\sqrt{2\pi N}} \sum_{k,n} \nonumber\\ &\times \left[ \int_{-\infty}^{0} \frac{{\rm d}p}{\sqrt{\omega_p}}  \, \int_{-\infty}^{+\infty} {\rm d}x\,  g_n(x) {\rm e}^{-i \left( p x - \frac{k x_n}{d} \right)}  l^{\dagger}(p) a_k  \right. \nonumber\\ &+  \left.  \int_{0}^{+\infty} \frac{{\rm d}p}{\sqrt{\omega_p}}   \, \int_{-\infty}^{+\infty} {\rm d}x\, g_n(x)  {\rm e}^{-i \left( p x - \frac{k x_n}{d} \right)} r^{\dagger}(p) a_k  \right. \nonumber\\ &- \left. {\rm h.c.} \right. \Big]
\end{align}
The TL only couples to the system at the position of the cavities, therefore, $g_n(x) = g \delta(x - x_n)$ and we have
\begin{align}\label{A7}
H_{\rm int} &= \frac{ivg}{\sqrt{2\pi N}} \sum_k \left( \int_{-\infty}^{0} \frac{{\rm d}p}{\sqrt{\omega_p}}  \,\sum_n  \left( l^{\dagger}(p) a_k {\rm e}^{-i\left(\frac{\omega_p}{v} + \frac{k}{d} \right) x_n} \right) \right. \nonumber\\  &+ \left. \int_{0}^{+\infty}  \frac{{\rm d}p}{\sqrt{\omega_p}} \,\sum_n \left( r^{\dagger}(p) a_k {\rm e}^{-i\left( \frac{\omega_p}{v} - \frac{k}{d} \right)x_n} \right) - {\rm h.c.} \right)
\end{align}
where we have replaced $p$ in the exponentials for $\pm \omega_p /v$. 
\begin{figure}[t]
\centering
\includegraphics[width=\columnwidth]{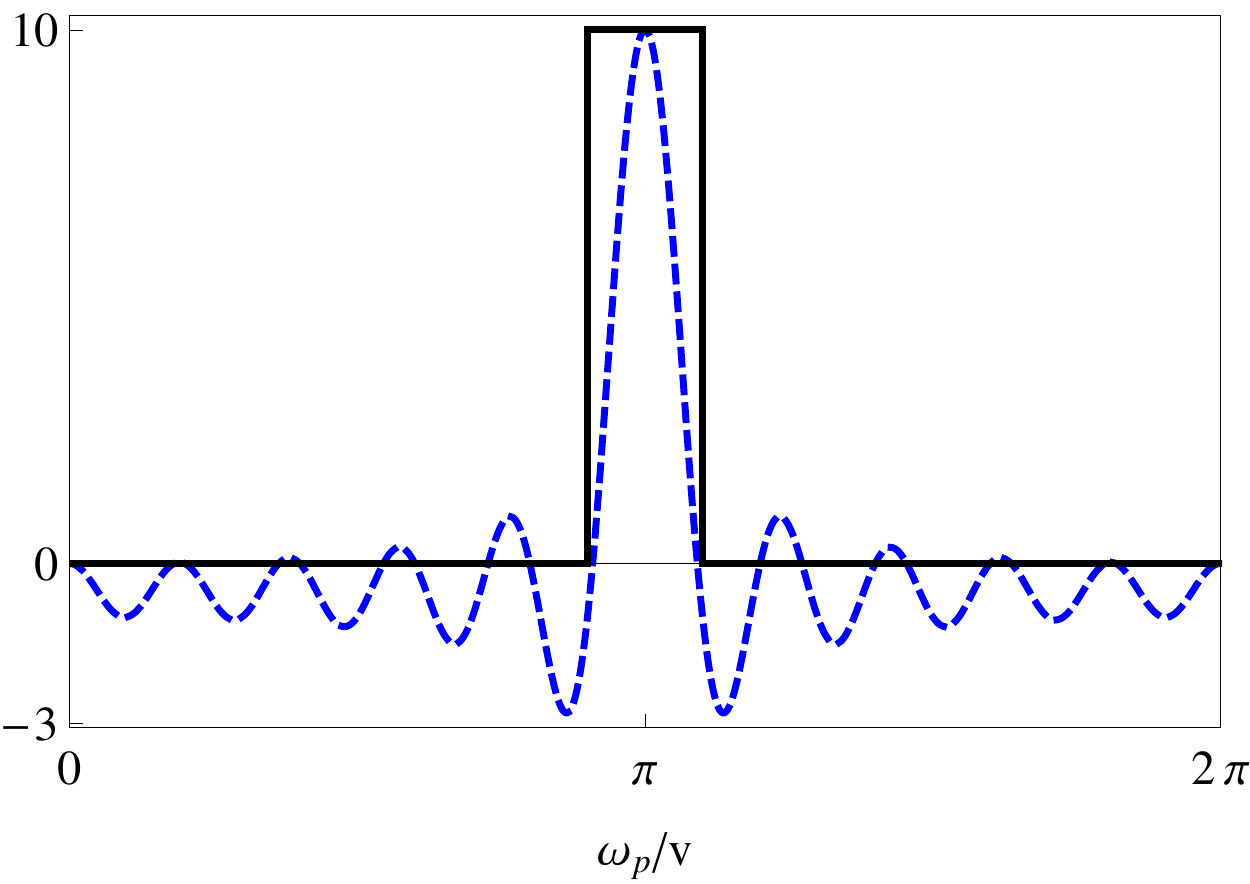}
\caption{(Color online)  Sum (\ref{rectangle}) (dashed) as a function of $\omega / d$ for $N=10$, and $k/d = \pi$.  Rectangle approximation of (\ref{rectangle}) (full line). This considers that the function is different from zero only in the region between its first two-zeros (symmetric around $k/d$) and that the height is constant and equal to $N$.}
\label{fig:dav}
\end{figure}
We can approximate the sum
\begin{equation}\label{rectangle}
\sum_n {\rm e}^{-i (\omega_p/v -k/d) x_n}
\end{equation}
by a rectangle of height $N$ and width $2\delta = 2\pi /N$ centered at $k$ (being zero elsewhere) (see Fig. \ref{fig:dav}). Similarly, $\sum_n {\rm e}^{i (\omega_p/v + k/d) x_n}$ will be replaced by a rectangle centered in $-k$. 
Therefore, we can rewrite (\ref{A7}) as 
\begin{eqnarray}\label{A8}
H_{\rm int} \sim \frac{ivg}{\sqrt{2\pi N}} \left( \int_{+k} \frac{{\rm d}p}{\sqrt{\omega_p}}\, N  \left( r^{\dagger}(p) a_k  - {\rm h.c.} \right) \right. \nonumber\\
\left. + \int_{-k} \frac{{\rm d}p}{\sqrt{\omega_p}} \, N  \left( l^{\dagger}(p) a_k  - {\rm h.c.} \right) \right),
\end{eqnarray}
where the integration intervals, following our previous approximation, are 
\begin{equation}
\label{A8b}
 \pm k \equiv\left[(\pm k-\delta)/d,( \pm k+\delta)/d\right].
\end{equation}

We are now able to write the Heisenberg equation of motion for the EM field operators. Following (\ref{A1}), (\ref{A2}) and (\ref{A8}) they are 
\begin{eqnarray}\label{A9}
\dot{r}(p) = -i vp \,r(p) + vg\sqrt{\frac{N}{2\pi \omega_p}} a_k(t) 
\end{eqnarray}
\begin{eqnarray}\label{A10}
\dot{l}(p) = +i vp \,l(p) + vg\sqrt{\frac{N}{2\pi \omega_p}} a_k(t)
\end{eqnarray}
Here we have included the explicit time dependence of the momentum cavity operators $a_k$. 
Integrating (\ref{A9}) from $t_0$ to $t$ ($t_0<t$) yields
\begin{eqnarray}\label{A11}
r(p,t) &=& {\rm e}^{-ivp(t-t_0)}r_0(p) \nonumber\\ &+& vg\sqrt{\frac{N}{2\pi \omega_p}} \int_{t_0}^t dt' e^{-ivp(t-t')} a_k(t')
\end{eqnarray}

\begin{figure}[t]
\centering
\includegraphics[width=\columnwidth]{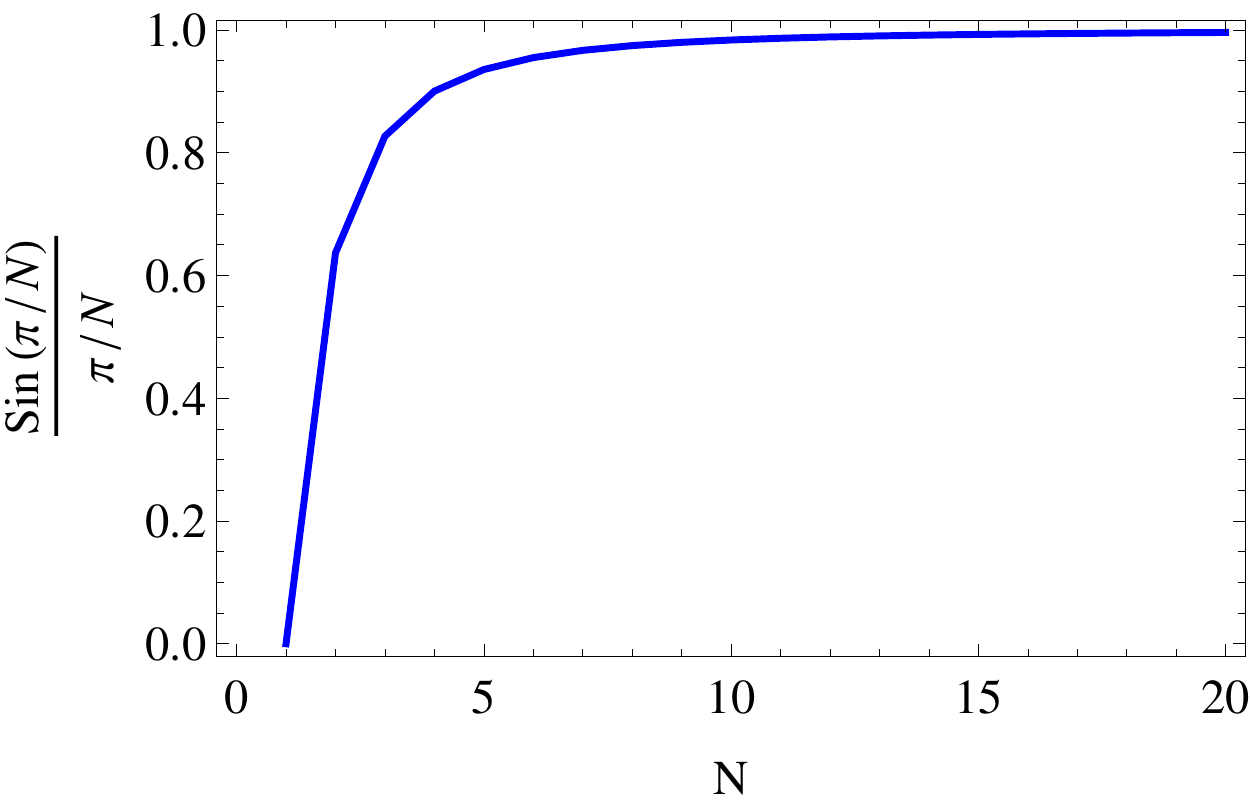}
\caption{(Color online) $\frac{\sin(y\pi/N)}{y\pi/N} $ as a function of the number $N$ of lattice sites. For $y \sim 1$ the limit $\lim_{z \rightarrow 0} \sin(z)/z = 1$ is very well approximated with a few number of sites ($N \sim 10$). }
\label{fig:approx2}
\end{figure}

where $r_0$ denotes the $r$ operator at time $t=t_0$. Notice that the former equations include a continuous momentum $p$ and a discrete one $k$. Recall from our approximation to the sum (\ref{rectangle}) that for any given $k$ only the momenta $p$ in a very narrow region of width $\delta$ (centered in $k$) contribute to our expressions. We now integrate (\ref{A11}) over this momentum interval ($+k$ for the right operators)
\begin{align}\label{A12}
\int {\rm d}p\, r(p,t) &= \sqrt{2\pi}r^{+k}_{in}(t) \nonumber\\ &+ g \sqrt{\frac{N v}{2\pi k}}  \int_{t_0}^t dt' \left( \frac{2 }{v(t-t')}  e^{-i kv(t-t')/d} \right. \nonumber\\ &\times \left. \sin \left(\frac{\delta v(t-t')}{d} \right) a_k(t') \right)
\end{align}
with the {\it input} operator $r^{+k}_{\rm in}$ defined as (following \cite{Gardiner1985})

\begin{equation}\label{A13}
r^{+k}_{\rm in}(t) = \frac{1}{\sqrt{2\pi}} \int_{(k-\delta)/d}^{(k+\delta)/d}{\rm d}p\, {\rm e}^{-ivp(t-t_0)} r_0(p)
\end{equation}

We have included the super index $+k$ to stress that this operator sums the momentum contributions in a narrow band around $k$ (positive for the right operators).
The $\emph{input}$ operator takes into account the free evolution of all  right-propagating EM  field modes before the interaction with the system. Therefore, it acts as a driving field in the equations of motion of the cavity operators. In the continuum limit ($\delta \rightarrow 0$) and for $t_0 \rightarrow -\infty$, (\ref{A12}) yields 
\begin{align}\label{C9}
\int {\rm d}p\, r(p,t) &= \sqrt{2\pi} \left( r^{+k}_{in}(t) +  \frac{g}{d} \sqrt{\frac{\pi v}{2 N k}} {\rm e}^{-ikvt/d} a_k \right)
\end{align}
as $\delta \rightarrow 0$, we make use of the following limit: $\lim_{z \rightarrow 0} \sin(z)/z = 1$ for $z = \delta v (t-t') /d$. This holds reasonably well for $N \gtrsim 10$ as it is shown in Fig. \ref{fig:approx2}. In addition, we have introduced
\begin{equation}\label{Ftime}
a_k =\frac{1}{\sqrt{2\pi}} \int_{-\infty}^{+\infty} {\rm d}t' {\rm e}^{ikvt'/d} a_k(t')
\end{equation}
In a similar way, we can integrate (\ref{A9}) from $t$ to $t_1$ ($t<t_1$) and define a corresponding \textit{output} operator. 
\begin{eqnarray}
r^{+k}_{\rm out}(t) = \frac{1}{\sqrt{2\pi}} \int_{+k} {\rm d}p \, {\rm e}^{-i vp(t-t_1)} r_1(p)
\end{eqnarray}
We will find that the \textit{input} and \textit{output} operators are related by
\begin{eqnarray}\label{Right}
r^{+k}_{\rm out}(t) = r^{+k}_{\rm in}(t) + \frac{1}{d} \sqrt{\frac{\Gamma v}{N k}} {\rm e}^{-ikvt/d} a_k
\end{eqnarray}
while for the left operators we find
\begin{eqnarray}\label{Left}
l^{-k}_{\rm out}(t) = l^{-k}_{\rm in}(t) + \frac{1}{d} \sqrt{\frac{\Gamma v}{N k}} {\rm e}^{+ikvt/d} a_k
\end{eqnarray}
where we have chosen conveniently $g = \sqrt{\Gamma/2\pi}$.

\end{appendix}

\bibliographystyle{apsrev4-1}
\bibliography{disolbib}

\end{document}